\documentstyle[epsfig]{aipproc}
\def\agt{
\mathrel{\raise.3ex\hbox{$>$}\mkern-14mu\lower0.6ex\hbox{$\sim$}}
}

\begin{document}
\title{Energetic Quantum Limit in Large-Scale Interferometers}


\author{Vladimir B.\ Braginsky, Mikhail L.\ Gorodetsky,$^*$ \\
Farid Ya.\ Khalili,$^*$ and Kip S.\ Thorne$^{\dagger}$}


\address{$^*$Physics Faculty, Moscow University, Moscow Russia \\
$^{\dagger}$Theoretical Astrophysics, California Institute of Technology,
Pasadena, CA 91125}

\maketitle

\abstract{  For each optical topology of an interferometric gravitational
wave detector, quantum mechanics dictates a minimum optical power (the
``energetic quantum limit'') to achieve a given sensitivity.  For standard
topologies, when one seeks to beat the standard quantum limit by
a substantial factor, the energetic quantum limit becomes impossibly large.
Intracavity readout schemes may do so with manageable optical powers.} 

\section{The Energetic Quantum Limit}

It is well known that quantum mechanics limits the sensitivity of traditional
position measurements by the Standard Quantum Limit (SQL).  Several
methods of overcoming the SQL have been 
proposed. It is likely that in
the next decade large-scale gravitational wave antennae will reach the
level of the SQL and possibly will beat it by factor $2\div 3$. Are there any
other quantum limits beyond the SQL?

One possible answer is: the next serious limitation is the Energetic
Quantum Limit. 
A gravitational wave in an interferometric antenna changes the
phase of the optical field. In order to detect this phase shift, the
uncertainty of
the phase $\Delta\phi$ must be sufficiently small. In particular, due to the
uncertainty relation

\begin{equation}
  \Delta{\cal E} \Delta\phi \ge \frac{\hbar\omega_0}{2}
\end{equation}
(where $\omega_0$ is the optical frequency), 
a large uncertainty of the optical energy $\cal E$ is required.

This is not a peculiar property of interferometric meters only, but a
consequence of a more general principle: In order to detect an external action
on a quantum object, the uncertainty of the interaction 
Hamiltonian $\hat{\cal
H}_I$ must be sufficiently large \cite{VMU83,VMU85,TheBook9}:

\begin{equation}
  \left\langle\left(\int_{-\infty}^{\infty}{\Delta\cal H}_I(t)dt\right)^2
  \right\rangle \ge \frac{\hbar^2}{4}\;.
  \label{EQL_1}
\end{equation}
The uncertainty of ${\cal H}_I$ is related to the signal-to-noise ratio by
\begin{equation}
  \frac{S}{N} = \frac{4}{\hbar^2}\left\langle\left(
    \int_{-\infty}^{\infty}{\Delta\cal H}_I(t)dt\right)^2\right\rangle
\end{equation}
For laser interferometer gravitational-wave antennas this is equivalent to

\begin{equation}
  \frac{S}{N} = \frac{4}{\hbar^2L^2}\int_{-\infty}^{\infty}
     x_{signal}(t)x_{signal}(t')B_{{\cal E}}(t,t')dtdt',
  \label{EQL_t}
\end{equation}
where $L$ is the length of the arms of the antenna,

\begin{equation}
  x_{signal}(t) = \frac{Lh(t)}{2}
\end{equation}
is the effective change of $L$ caused by a gravitational wave, $h(t)$ is
the variation of the wave's metric, and $B_{{\cal E}}(t,t')$ is the
correlation function of the optical
energy in the antenna, or the correlation function of the difference of 
energies in the two arms of the antenna if two-arm-topology is used.

The origin of the limitation ({\ref{EQL_1}) 
is the Heisenberg uncertainty relation. To
detect a small displacement of the mirrors, it is necessary to apply to 
them a sufficiently
strong random kick. The only source of this kick is the uncertainty of the
optical energy in the antenna or of the difference of energies in the two arms.

We shall limit ourselves here to the stationary regime, 
for which the quantum state of the
electromagnetic field in the interferometer does not depend explicitly on time.
In this case formula (\ref{EQL_t}) can be rewritten in spectral
form as

\begin{equation}
  \frac{S}{N} = \frac{4}{\hbar^2L^2}\int_{-\infty}^{\infty}
    |X_{signal}(\omega)|^2 S_{{\cal E}}(\omega) \frac{d\omega}{2\pi}\;,
  \label{EQL_f}
\end{equation}
where $X_{signal}(\omega)$
%
is the Fourier transform of $x_{signal}(t)$, and $S_{{\cal E}}(\omega)$
%
is the spectral density of the fluctuations of the optical energy. It is
important to note here that formula (\ref{EQL_t}) is the ultimate limit
on the sensitivity for any measurement technique, and formula (\ref{EQL_f})
describes the ultimate sensitivity for all stationary procedures.

\section{Comparison with the SQL}

In all estimates below we will use the value of the Standard Quantum Limit
(SQL) as a convenient measure of sensitivity. 
The SQL, as it was defined more than
thirty years ago \cite{Braginsky67}, is the sensitivity of an ordinary position
meter, i.e.\ a position meter which does not use any non-stationary or
correlation methods to increase the sensitivity. The forms of the SQL as usually
given in the literature, 
are not convenient since they are based on some assumed
shape of the force's time dependence (most commonly a single-cycle sinusoid
or a long, monochromatic wave train). Here we prefer a more general form of
the SQL expressed in terms of the spectral density $S(\omega)$ for the net 
noise of a measurement device. This (double-sided)
spectral density is defined in such a way that
for optimal signal processing the signal to noise ratio is equal to

\begin{equation}
  \frac{S}{N} = \int_{-\infty}^\infty
    \frac{|X_{signal}(\omega)|^2}{S(\omega)}\frac{d\omega}{2\pi}.
  \label{S/N}
\end{equation}
In the case of an ordinary position meter the spectral density of the net
noise is equal to

\begin{equation}
  S_{SQL}(\omega)=S_x + \frac{S_F}{M^2\omega^4}, \label{SSxSf}
\end{equation}
where $M$ is the test mass, $S_x(\omega)$ is the spectral density of the
noise that the meter superimposes on the output position signal,
$S_F(\omega)$ is the spectral density of the fluctuating back-action force
that the meter exerts on the test mass, and these noises satisfy 
the uncertainty
relation \cite{TheBook6},

\begin{equation}
  S_x S_F \ge \hbar^2/4 \;.
\end{equation}
We assume that the position meter is a perfect one, corresponding to equality
in this uncertainty relation, and the spectrum of the signal is concentrated
near the frequency $\omega_{signal}$. In this case the meter can be optimally
tuned to minimize the net noise at this frequency:

\begin{equation}
  S_{SQL}(\omega_{signal}) = \frac{\hbar}{M\omega_{signal}^2}. \label{SQL}
\end{equation}
As follows from formulas (\ref{EQL_f}), (\ref{S/N}) and (\ref{SQL}), in
order to beat this SQL for the spectral density of $x_{signal}$ by a factor 
$\xi^2 < 1$, the spectral
density of the fluctuations of the optical energy must obey the condition

\begin{equation}
  S_{{\cal E}}(\omega_{signal}) \ge
  \frac{\hbar^2}{S_{SQL}(\omega_{signal})\xi^2} =
  \frac{\hbar M\omega_{signal}^2L^2}{4\xi^2}. \label{EQL}
\end{equation}
Let us calculate now the values of $S_{{\cal E}}$ for different possible
topologies of gravitational-wave antennae.

\begin{figure}[b!] 
\centerline{\epsfxsize=3.5in\epsfbox{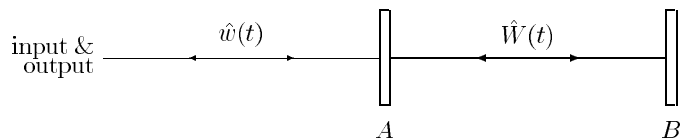}}
\vspace{10pt}
\caption{One-arm topology.}
\label{fig1}
\end{figure}

\section{One-arm scheme}

The simplest of possible topologies, shown in Fig.\ \ref{fig1}, 
consists of
a single Fabry-Perot resonator excited by a pumping laser. The optimally chosen
variable of the reflected beam is registered. 
A spectral or time-domain variational
measurement \cite{VarMeas}, or any other advanced technique can be used to
increase the sensitivity. Simple calculations yield that in this case

\begin{equation}
  S_{{\cal E}}(\omega) =
  \frac{2\zeta^2\hbar\omega_0\overline{{\cal E}}\delta}{\delta^2+\omega^2},
  \label{S_naive}
\end{equation}
where $\omega_0$ is the eigenfrequency of the resonator, $\delta$ is the
half-bandwidth of the resonator, $\overline{{\cal E}}$ is the mean value of
the optical energy in the resonator, and $\zeta$ is the squeeze factor of the
quantum state of the input wave ($\zeta=1$ for a coherent quantum state 
and $\zeta>1$
for a squeezed state). As was noted by Caves almost twenty years ago
\cite{Caves80}, the use of a squeezed state allows one to reduce the value of
$\overline{{\cal E}}$ by the factor $\zeta^2$. Unfortunately, experimental
achievements in squeezing are still quite modest, so we will use $\zeta=1$ 
(i.e.\ coherent state) below in all formulas and numerical estimates.

Substituting formula (\ref{S_naive}) and $\zeta=1$
into (\ref{EQL}), we obtain that in this case

\begin{equation}
  \overline{{\cal E}} =
  \frac{M\omega_{signal}^2(\delta^2+\omega_{signal}^2)L^2}
    {8\xi^2\omega_0\delta}\;,
  \label{E_naive}
\end{equation}
and, correspondingly,

\begin{equation}
  \overline{W} = \frac{M\omega_{signal}^2(\delta^2+\omega_{signal}^2)Lc}
    {16\xi^2\omega_0\delta}\;,
  \label{W_naive}
\end{equation}

\begin{equation}
  \overline{w} = \frac{M\omega_{signal}^2(\delta^2+\omega_{signal}^2)L^2}
    {16\xi^2\omega_0}\;,
  \label{w_naive}
\end{equation}
where $\overline{W}$ is the mean value of the optical power circulating in the
resonator and $\overline{w}$ is the mean value of the input pump power.

The values of $\overline{{\cal E}}$ and $\overline{W}$ may be minimized by
choosing for the resonator's half-bandwidth

\begin{equation}
  \delta=\omega_{signal} \label{delta_opt}\;.
\end{equation}
In this case

\begin{eqnarray}
  \overline{{\cal E}} &=& 
  \frac{M\omega_{signal}^3L^2}{4\xi^2\omega_0} \nonumber\\
  &=& \frac{2\cdot 10^8 {\rm erg}}{\xi^2}
    \left(\frac{M}{10^4{\rm g}}\right) 
    \left(\frac{\omega_{signal}}{10^3{\rm s}^{-1}}\right)^3 
    \left(\frac{L}{4\cdot 10^5{\rm cm}}\right)^2 
    \left(\frac{\omega_0}{2\cdot 10^{15}{\rm s}^{-1}}\right)^{-1}, \nonumber\\
  \label{E_naive_opt}
\end{eqnarray}

\begin{eqnarray}
\overline{W} &=& \frac{M\omega_{signal}^3Lc}{8\xi^2\omega_0} 
\nonumber\\
 &=& \frac{0.75\cdot 10^{13} {\rm erg/s}}{\xi^2} 
    \left(\frac{M}{10^4{\rm g}}\right) 
    \left(\frac{\omega_{signal}}{10^3{\rm s}^{-1}}\right)^3 
    \left(\frac{L}{4\cdot 10^5{\rm cm}}\right) 
    \left(\frac{\omega_0}{2\cdot 10^{15}{\rm s}^{-1}}\right)^{-1} ,
\nonumber\\
  \label{W_naive_opt}
\end{eqnarray}

\begin{eqnarray}
  \overline{w} &=& \frac{M\omega_{signal}^4L^2}{8\xi^2\omega_0} 
\nonumber\\
  &=& \frac{1\cdot 10^{11} {\rm erg/s}}{\xi^2} 
    \left(\frac{M}{10^4{\rm g}}\right) 
    \left(\frac{\omega_{signal}}{10^3{\rm s}^{-1}}\right)^4 
    \left(\frac{L}{4\cdot 10^5{\rm cm}}\right)^2 
    \left(\frac{\omega_0}{2\cdot 10^{15}{\rm s}^{-1}}\right)^{-1} .
\nonumber\\
  \label{w_naive_opt}
\end{eqnarray}

All known methods of overcoming the SQL for position measurements
of free test masses \cite{VarMeas,Unruh,speedmeter} apply an additional 
restriction on
the optical energy and pumping power.  This restriction arises as follows. 
All such methods rely on
a correlation between two meter noises: the (radiation-pressure-induced) 
back action noise on the test masses, 
and the output light's shot noise.  This correlation is damaged by
electromagnetic fluctuations which enter the resonator 
wherever there is optical
dissipation.  Such fluctuations produce
radiation pressure noise that cannot correlate with the shot noise, so
all these schemes are limited by the level of intrinsic
losses (dissipation) in the electromagnetic resonator.  It can be shown
that, due to this, the amount by which these methods beat the SQL cannot
exceed the limit

\begin{equation}
  \xi = \left(\frac{\delta_{intr}}{\delta_{load}}\right)^{1/4}, \label{5_1}
\end{equation}
where $\delta_{intr}$ is the part of the electromagnetic
resonator's half-bandwidth
$\delta=\delta_{intr}+\delta_{load}$ due
to intrinsic losses (i.e. absorption in the
mirrors and beam-splitters and nonzero transmittance of end mirrors) and
$\delta_{load}$ is the part which characterizes the coupling of the resonator
to the output light beam. 
Hence the optimization (\ref{delta_opt}) is possible only if

\begin{equation}
  \xi > \left(\frac{\delta_{intr}}{\omega_{signal}}\right)^{1/4}\;.
\label{xideltaomega}
\end{equation}
Using the best known mirrors and the LIGO value of $L=4\times 10^5{\rm cm}$ 
one can
achieve the value of $\delta_{intr} \sim 10^{-1}{\rm s}^{-1}$. Hence, if
$\omega_{signal} \simeq 10^3{\rm s}^{-3}$ then condition (\ref{5_1}) may be
satisfied only if $\xi\ge 0.1$.

Any further increase of sensitivity (decrease of $\xi$)
makes the optimization (\ref{delta_opt})
impossible. Instead, the resonator's half-bandwidth will be forced to increase,
with decreasing $\xi$, as

\begin{equation}
  \delta \simeq \delta_{load} = \frac{\delta_{intr}}{\xi^4} > 
\omega_{signal}\;,
\end{equation}
which leads to the necessity of increasing the pumping power:

\begin{equation}
  \overline{{\cal E}} =
  \frac{M\omega_{signal}^2\delta_{intr}L^2}{8\xi^6\omega_0}\;,
  \label{E_naive_6}
\end{equation}

\begin{equation}
  \overline{W} =
  \frac{M\omega_{signal}^2\delta_{intr}Lc}{16\xi^6\omega_0\delta}
  \label{W_naive_6}\;,
\end{equation}

\begin{equation}
  \overline{w} =
  \frac{M\omega_{signal}^2\delta_{intr}^2L^2}{16\xi^{10}\omega_0}\; .
  \label{w_naive_10}
\end{equation}
Numerical values of ${\cal E}$, $\overline{W}$ and $\overline{w}$ calculated
using formulas (\ref{E_naive_opt})--(\ref{w_naive_opt}) and
(\ref{E_naive_6})--(\ref{w_naive_10}) are plotted in Figs.\
\ref{fig5}, \ref{fig6} and \ref{fig7} correspondingly,
as a functions of $\xi^{-1}$ (see the upper solid curves at the end of this
paper). 
In these figures, LIGO parameter values are used:
$M=10^4{\rm g}$, $L=4\times 10^5{\rm cm}$, $\omega_0=2\times
10^{15}{\rm s}^{-1}$, and $\omega_{signal}=10^3{\rm s}^{-1}$. 
These figures show a
very strong dependence of ${\cal E}$, $\overline{W}$ and $\overline{w}$ on
$\xi$ in the region to the left of the inflection point
$\xi=(\delta_{intr}/\omega_{signal})^{1/4}$.  

This strong $\xi$ dependence
($\overline{\cal E} \propto \xi^{-6}$, $\overline{W}\propto \xi^{-6}$,
$\overline{w} \propto \xi^{-10}$)
makes it practically impossible
to achieve sensitivities $\xi<0.1$ using conventional interferometer
topology:  For $\xi = 0.1$ the circulating power in the resonator arms
is $\overline{W} \sim 100 {\rm MW}$, corresponding to a power density 
$\sim 300 {\rm kW/cm}^2$ in the $\sim 10 {\rm cm}$-diameter light spots
on each mirror's surface.  This power density is already so high that it is
questionable whether future mirrors will be able to handle it; and the 
further increase as $\overline{W}\propto\xi^{-6}$ makes it highly implausible 
that future mirrors will allow sensitivites better than $\xi \simeq 0.1$.

\section{Recycling schemes}

In the next generation of large-scale gravitational-wave antennae, 
recycling \cite{Recycl} will be used for several purposes, including
reduction of the optical pump power
$\overline{w}$. Meers
\cite{Meers} has discussed several different variants of recycling, which 
can be divided into two main groups, depending on the structure of the 
electromagnetic modes
in the antenna: wide-band recyling, and narrow-band or resonant recycling. 
We shall
illustrate these two types of recycling schemes by two concrete topologies,
but the main properties in each group are independent of the concrete topology.

\subsection{Wide-band recycling}

In wide-band recycling schemes (Fig.\ \ref{fig2}) the optical field in the 
interferometer
may be regarded as a superposition of two electromagnetic 
modes -- the input mode (also
called ``symmetric mode''; solid
lines in Fig.\ \ref{fig2}) 
and the output mode (also called ``antisymmetric mode'';
dashed lines). The
Bandwidths of these modes
depend on the transmittances of the mirrors $D$ and $D'$, correspondingly,
and may be substantially different. Mirror $D'$ may be completely absent, 
which corresponds to the case of simple power recycling.

\begin{figure}[b!] 
\centerline{\epsfxsize=2.5in\epsfbox{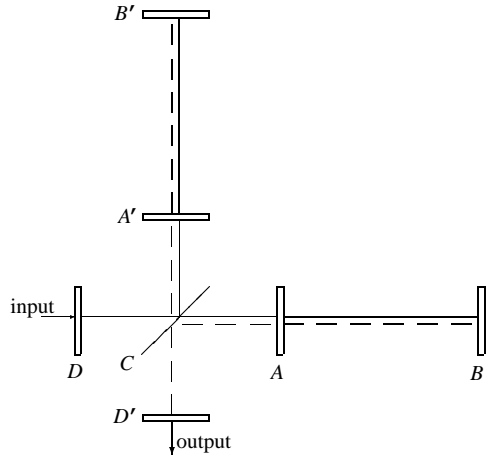}}
\vspace{10pt}
\caption{Wide-band recycling toplogy.}
\label{fig2}
\end{figure}

All previous expressions for the mean resonator energy $\overline{{\cal E}}$ and
circulating power $\overline{W}$ [formulas (\ref{E_naive}),
(\ref{W_naive}), (\ref{E_naive_opt}), (\ref{W_naive_opt}), (\ref{E_naive_6})]
remain valid for wide-band recycling schemes,
but with replacement of $\delta$ by the
half-bandwidth of the output mode $\delta_{out}$.  On the other hand, the
value of the input (pump) power is defined by the bandwidth of the input mode:

\begin{equation}
  \overline{w} = 2\overline{\cal E}\delta_{in} .
\end{equation}
This allows one to decrease the input power by holding $\delta_{out} \sim
\omega_{signal}$ and choosing $\delta_{in}$ as small as possible. For
example, using the best known mirrors, the value of $\delta_{in}\sim
10^{-1}{\rm s}^{-1}$ can be obtained, 
which allows one to decrease the input power by
several orders of magnitude (see the dotted curve in Fig.\ \ref{fig7}).

\subsection{Narrow-band recycling}

Narrow-band recycling schemes differ from wide-band ones only by
the values of the distances between mirrors and the mirror transmittances. 
However, narrow-band schemes lead to a
substantially different behavior of the system. Its eigenfrequencies form
doublets. The frequencies in each doublet are separated from each other by 
a frequency difference
$\Omega$ which depends on the the mirror configuration and which should be 
chosen close to $\omega_{signal}$.  The spectral density of stored energy
$S_{{\cal E}}$ in this case is maximimal at the frequency $\Omega$ instead 
of at zero frequency:

\begin{equation}
  S_{{\cal E}} = \frac{4\hbar\omega_0\overline{{\cal E}}\omega^2\delta}
    {(\Omega^2-\omega^2)^2+\Omega^2\delta^2} .
\end{equation}
This allows one to obtain an especially high value of $S_{{\cal E}}$ in a narrow
band in the vicinity of $\Omega$ by choosing $\Omega \gg \delta$. It can be
shown that in this case

\begin{equation}
  \overline{{\cal E}} =
  \frac{M\omega_{signal}^2(\delta^2+\Delta^2)L^2}{8\xi^2\omega_0\delta}
  \label{E_narrow}
\end{equation}
where $\Delta$ is the bandwidth over which $\xi$ is close to its desired
minimum value. This stored energy can be further minimized by choosing
$\delta=\Delta$, so that

\begin{equation}
  \overline{{\cal E}} =
  \frac{M\omega_{signal}^2\Delta L^2}{4\xi^2\omega_0}\; .
  \label{E_narrow_opt}
\end{equation}
It is evident from comparison of formulas (\ref{E_naive_opt}) and
(\ref{E_narrow_opt}) that this regime allows one to reduce the necessary 
value of
${\cal E}$, but the price is a proportional reduction of the bandwidth of
the signal.

\section{Intracavity schemes}

The above considerations show that
the Energetic Quantum Limit
does not permit one to obtain sensitivities substantially better than 
the SQL using the traditional optical topologies of gravitational-wave 
antennae, even if recycling is employed.

A new method of extracting information from a gravitational-wave
antenna was proposed in the article \cite{Intra96}. Instead of measuring the
time phase shift of an output optical wave, it was suggested to detect directly
the spatial phase shift of the optical field inside the antenna using some
QND-type method. In this case the necessary level of uncertainty of the optical
energy (\ref{EQL_1}) is provided by a fluctuational redistribution of the 
energy between the two arms of the antenna instead of by the shot noise of 
the resonator energy. In principle
this allows one to increase the sensitivity without increasing the mean 
stored energy.

Two practical realizations of this idea were considered in the articles
\cite{Intra97,Intra98}. From our point of view, the most promising of them
is the ``optical bar'' scheme \cite{Intra97}. The structure of the 
electromagnetic modes in this case
is similar to that of narrow-band recycling topologies: It has frequency
doublets separated by a frequency $\Omega$, which depends on 
the transmittance
of the central mirror.  One of the possible topologies for the ``optical bar''
scheme is shown in Fig.\ \ref{fig3}.

\begin{figure}[b!] 
\centerline{\epsfxsize=1.9in\epsfbox{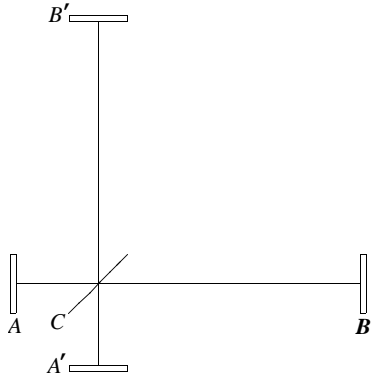}}
\vspace{10pt}
\caption{Optical bars.}
\label{fig3}
\end{figure}

In this scheme (we omit now all intermediate details) the optical fields in 
the two
arms of the antenna work (via their radiation pressure)
as mechanical springs with rigidity proportional
to the optical energies $\overline{{\cal E}}$ stored in 
the two arms.  The optical fields' rigidities move
the internal mirror $C$ when a gravitational wave moves the end mirrors. This
displacement may be measured relative to an additional reference mass which
is not affected by the optical field in the antenna, using one of the
measurement methods developed for resonant-mass gravitational-wave antennae.

The value of the displacement of mirror $C$ may be close to $Lh(t)/2$ if 
the stored energy
$\overline{{\cal E}}$ is sufficiently high:

\begin{equation}
  \overline{{\cal E}} \ge \frac{m\omega_{signal}^3L^2}{2\omega_0}\;,
  \label{E_intra}
\end{equation}
where $m$ is the mass of the
internal mirror.  The structure of this expression is
similar to expression (\ref{E_naive_opt}) for traditional schemes, but it
contains the mass $m$ of the internal mirror $C$ instead of the masses of 
the end
mirrors $M$, and, which is more important, it does not depend on $\xi$. Hence,
in this case it is possible to increase the sensitivity beyond the level of 
the SQL
without any necessity to increase $\overline{{\cal E}}$. If, for example, 
$m=10^3 {\rm g}$
and the values of all other parameters are the same as above, then
$\overline{{\cal E}} \simeq 4\cdot 10^7 {\rm erg}$.

It should be noted that internal losses in the interferometer limit the
sensitivity of the intracavity schemes at the level,

\begin{equation}
  \xi \simeq \sqrt\frac{\delta_{intr}}{\omega_{signal}}
\end{equation}
[square root, 
by contrast with the $1/4$ power for conventional interferometers, Eq.\ 
(\ref{xideltaomega})];
so if $\delta_{intr}= 10^{-1} {\rm s}^{-1}$, then the value 
$\xi \simeq 10^{-2}$ may be reached
in principle.

\section{Methods of measurement}.

It is evident that for intracavity schemes the focus of the problem
shifts to the local meter that measures the position of the central mirror. 
Intracavity schemes require high precision in this meter's measurement, 
exceeding the
SQL. During the last twenty five years several possible methods of such 
measurement were proposed. We consider here briefly only one of them -- the
so-called `` dual-resonator speedmeter'' \cite{speedmeter,speedmeter2}. It 
is based on two
microwave resonators, one of which couples evanescently to the position of the
test mass (see Fig.\ \ref{fig4}). The sloshing of the resulting signal between the 
resonators, and a
wise choice of where to place the resonators' output waveguide, produce a
signal in the waveguide that (for sufficiently low frequencies) is
proportional to the test-mass velocity, but not its position. This permits
the speed meter to achieve sensitivities better than the SQL.

\begin{figure}[b!] 
\centerline{\epsfxsize=2.5in\epsfbox{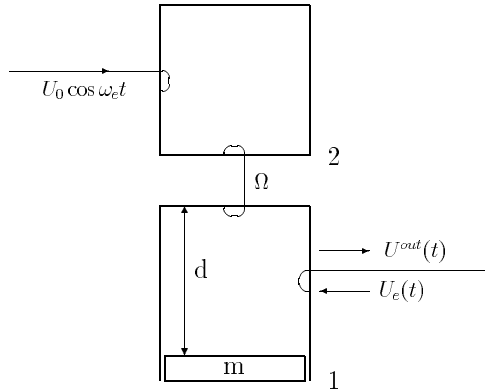}}
\vspace{10pt}
\caption{Quantum speed meter.}
\label{fig4}
\end{figure}

Unfortunately, this scheme has two major disadvantages that are typical for all
similar schemes. First, when using this scheme to monitor the central mass,
the level of optical energy in the antenna required
to overcome the SQL has to be substantially higher than the theoretical
limit (\ref{E_intra}):

\begin{equation}
  \overline{{\cal E}} \ge \frac{M\omega_{signal}^3L^2}{2\xi^2\omega_0}
  \label{E_speed}
\end{equation}
and even a bit higher than the energy (\ref{E_naive_opt}) (see 
the dotted curves in
Figs.\ \ref{fig5}, \ref{fig6}, \ref{fig7}. This is because 
the speedmeter satisfies the criterion for QND 
measurements only when monitoring a 
free mass, and the internal mirror $C$ coupled to the optical field and through
this field to the end mirrors is a more sophisticated object than a free mass;
it has several degrees of freedom.  This mirror behaves like a
single solid free mass only if the optical energy is sufficiently 
large, which leads to the requirement (\ref{E_speed}).

The second disadvantage is an inherent property of all known devices for 
overcoming
the SQL for position measurements of free test masses.\  As 
mentioned above, all such measurement devices are very sensitive to
the level of intrinsic losses in their resonators [see formula (\ref{5_1}) and
the discussion above]. In the particular case of the speedmeter, this 
disadvantage limits its sensitivity to $\xi \agt 1/3$ even if the best known
microwave resonators are used \cite{speedmeter2}.  An additional fee for this
scheme is the necessity to use cryogenic temperatures.

The first disadvantage (abnormally high optical power)
may be evaded by using a local ``variational meter'' \cite{VarMeas} based
on a microwave parametric transducer or on a relatively short Fabry-Perot 
sensor.
As follows from the formula (\ref{W_naive_opt}), the necessary 
circulating optical power is proportional to the length of the resonator.
But for such a variational meter, 
the second disadvantage (a severe limitation on $\xi$ due to 
electromagnetic dissipation) still remains.

Two co-authors of this report think that the most promising way to
evade this obstacle is to use an oscillator instead of free test mass \cite{Rigidity}. 
In the case of a 
harmonic oscillator it is possible to overcome the SQL without using
a correlation of meter noises, and thus the oscillator is free from the 
limitation (\ref{5_1}).

\begin{figure}[b!] 
\centerline{\epsfxsize=3.5in\epsfbox{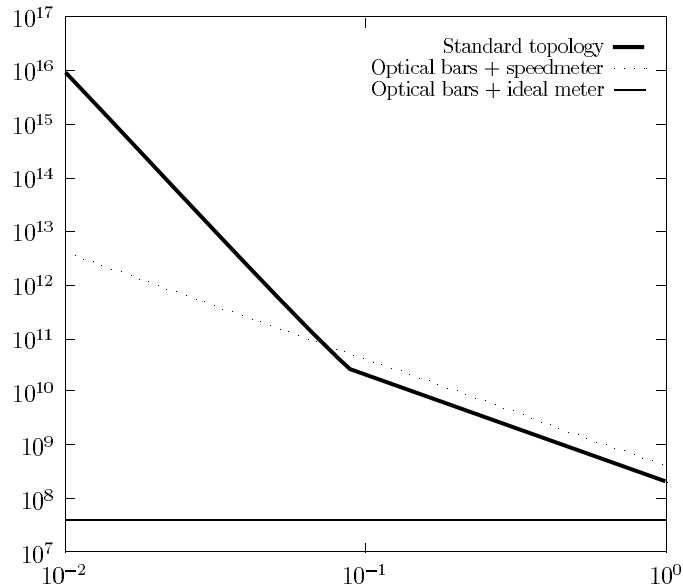}}
\vspace{10pt}
\caption{Energy as a function of $\xi$, erg.}
\label{fig5}
\end{figure}

\begin{figure}[b!] 
\centerline{\epsfxsize=3.5in\epsfbox{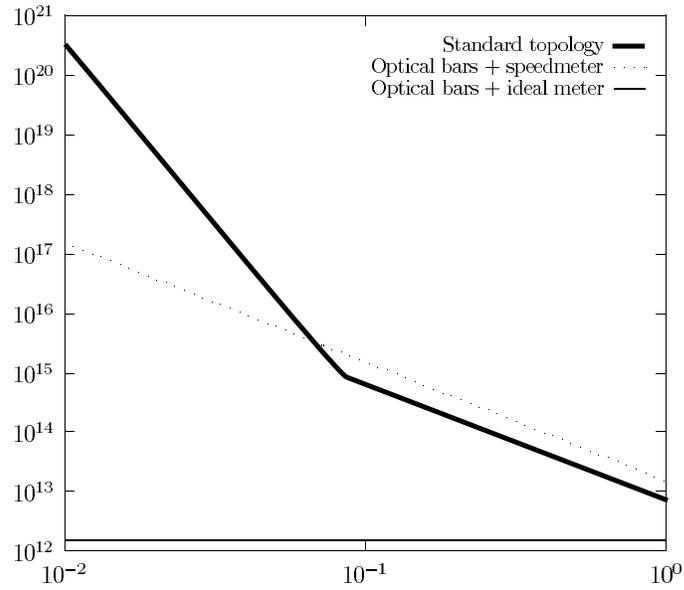}}
\caption{Circulating power as a function of $\xi$, erg/s.}
\label{fig6}
\end{figure}

\begin{figure}[b!] 
\centerline{\epsfxsize=3.5in\epsfbox{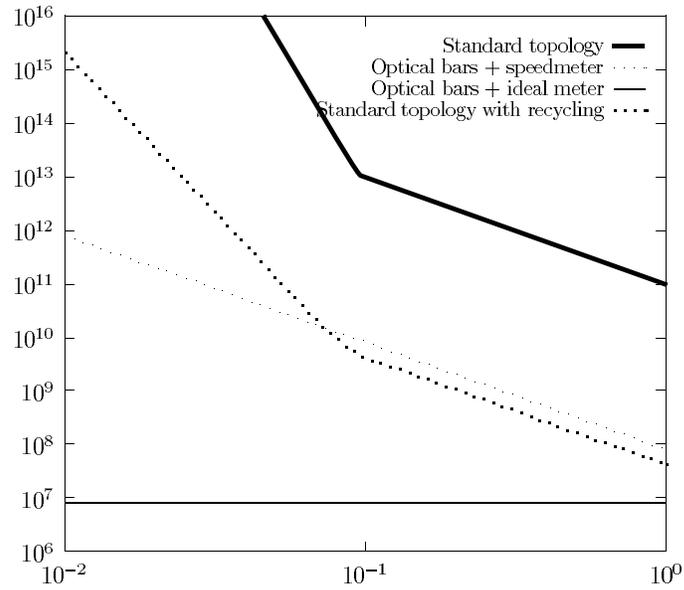}}
\vspace{10pt}
\caption{Pump power as a function of $\xi$, erg/s.}
\label{fig7}
\end{figure}

\end{document}